\documentclass[fdp,a4paper,fleqn%
]{w-art}
\usepackage{times,cite,w-thm}

\usepackage{amsthm,amssymb,amsmath,bm,mathrsfs}

\theoremstyle{plain}

\theoremstyle{definition}

\usepackage[]{graphicx}

\newcommand{\be}{\begin{equation}}
\newcommand{\ee}{\end{equation}}
\newcommand{\ba}{\begin{eqnarray}}
\newcommand{\ea}{\end{eqnarray}}

\begin{document}
\DOIsuffix{theDOIsuffix}
\pagespan{1}{}



\title[Comments on higher-spin holography]{Comments on higher-spin holography}


\author[X. Bekaert]{Xavier Bekaert\inst{1,}%
  \footnote{E-mail:~\textsf{xavier.bekaert@lmpt.univ-tours.fr}}}
\address[\inst{1}]{Laboratoire de Math\'ematiques et Physique Th\'eorique\\
Unit\'e Mixte de Recherche $7350$ du CNRS, F\'ed\'eration de Recherche $2964$ Denis Poisson\\
Universit\'e Fran\c{c}ois Rabelais, Parc de Grandmont, 37200 Tours, FRANCE}
\author[E. Joung]{Euihun Joung\inst{2,}\footnote{E-mail:~\textsf{euihun.joung@sns.it}}}
\address[\inst{2}]{Scuola Normale Superiore and Istituto Nazionale di Fisica Nucleare\\
Piazza dei Cavalieri 7, I-56126 Pisa, ITALY}
\author[J. Mourad]{Jihad Mourad\inst{3,}\footnote{E-mail:~\textsf{mourad@apc.univ-paris7.fr}}}
\address[\inst{3}]{AstroParticule et Cosmologie\\
Unit\'e Mixte de Recherche $7164$ du CNRS\\
Universit\'e Paris VII, B\^atiment Condorcet, 75205 Paris Cedex 13, FRANCE}

\begin{abstract}
The conjectured holographic duality between vector models with quartic interaction and higher-spin field theory in the bulk is reviewed, with emphasis on some versions and generalisations (higher dimensions, beyond the singlet sector, etc) which have not been much investigated yet.
The strongest form of the conjecture assumes that it holds for any (not necessarily large) number of massless scalar fields and for any value of the coupling constant.
Since the quartic interaction is of double-trace type, the exact duality (for any value of the coupling constant) automatically follows from its validity at the Gaussian fixed point (for vanishing coupling constant). The validity of the latter also implies that unbroken higher spin symmetries should prevent quantum corrections in the bulk.
\end{abstract}
\maketitle                   





\section{Introduction}

The AdS/CFT correspondence and the higher-spin field theory appear closely intertwined. The exploration of
their profound connections might provide a key towards a deeper understanding of their mysterious properties.
For instance, the oldest precursor of the AdS/CFT correspondence is presumably
the Flato-Fronsdal theorem \cite{Flato:1978qz} providing the description of elementary gauge fields living in the interior of $AdS_4$ in terms of two singletons, \textit{i.e.} elementary massless fields living on the conformal boundary of $AdS_4$.
This theorem prompted Fronsdal to pursue his seminal study of higher-spin gauge fields on anti de Sitter spacetime \cite{Fronsdal:1978vb}, due to the appearance of an infinite tower of gauge fields with unbounded spin in the decomposition of the product of two singletons.
Although the physical interpretation of the theorem has changed\footnote{The old interpretation was that bulk gauge fields are composite fields made of two singletons, while the modern interpretation is rather that bulk gauge fields \textit{couple}, through their boundary data, to composite fields (now interpreted as boundary bilinear currents) made of two singletons. Of course, this subtler relation is a crucial conceptual ingredient in the AdS/CFT correspondence.}, it remains instrumental for higher-spin symmetries, interactions, holography, \textit{etc}.

The precise connections between both subjects emerged progressively during the first years of this century.
Initiated on $AdS_5\times S^5$ in the context of Maldacena's conjecture \cite{Sundborg:2000wp},
these ideas were pursued in any spacetime dimensions, first at the level of kinematics
\cite{Konstein:2000bi} and later at a dynamical level,
leading to the duality conjecture between bosonic higher-spin gravity around $AdS_{d+1}$ (for any $d>2$) and a $d$-dimensional conformal field theory (CFT$_d$) of free massless scalars in the vector representation of an internal symmetry
group \cite{Mikhailov:2002bp,Sezgin:2002rt}, refined to include the three-dimensional strongly-coupled
critical $O(N)$ and Gross-Neveu models respectively
in \cite{Klebanov:2002ja} and \cite{Sezgin:2003pt}.
The concrete relation with Vasiliev's unfolded equations in four and five dimensions\footnote{Notice that the complete non-linear equations of bosonic higher-spin gravity in any higher dimension were actually presented in \cite{Vasiliev:2003ev} after the formulation of the conjectures \cite{Mikhailov:2002bp,Sezgin:2002rt,Klebanov:2002ja,Sezgin:2003pt} so, strictly speaking, the conjecture in any dimension is a partially retrospective perspective.} was elaborated in
\cite{Sezgin:2002rt,Sezgin:2003pt}.
The recent checks of the conjecture for $AdS_4/CFT_3$ at cubic level \cite{Giombi:2009wh} prompted a revived interest in the correspondence.
For instance, the conjecture has been generalised in the presence
of a Chern-Simons gauge field on the three-dimensional boundary \cite{Aharony:2011jz} and another duality has been proposed relating bosonic Vasiliev's theory on de Sitter bulk spacetime $dS_4$
and fermionic scalar Euclidean $CFT_3$ \cite{Anninos:2011ui}.
Several roads toward a constructive derivation of the bulk dual of a free CFT have been proposed, such as the bilocal field  \cite{Jevicki:2011ss} and the renormalisation group \cite{Douglas:2010rc} approaches.

This short note intends to briefly review the strong form of the conjectured holographic duality between Vasiliev bosonic higher-spin gravity and the theory of massless scalar fields in the vector representation of the internal symmetry group with quartic interactions.
The goal is to underline few basic facts:

$\bullet$ Vasiliev bosonic higher-spin gravity exists in explicit form at tree level in any spacetime dimension and for any (classical compact semi-simple Lie) group as internal symmetry, at the level of complete field equations \cite{Vasiliev:2003ev}.\footnote{
More precisely, the bulk gauge fields (and thus the boundary bilinear currents) can be matrix-valued: in the adjoint (or anti/\,symmetric) representation of the unitary (or orthogonal/symplectic) groups; see the conclusion of \cite{Vasiliev:2003ev} for more details.}
Therefore, the conjecture can be defined in any spacetime dimension and is not necessarily restricted to the sector of singlets of the internal symmetry group.\footnote{Both generalisations play an important role in the proposal \cite{Bekaert:2011cu} of a gravity dual of the non-relativisitc Fermi gas at unitarity.}

$\bullet$ At the dynamical level (\textit{i.e.} beyond two-point functions), the relevance of the quartic interaction as well as the stability or unitarity of the interacting boundary theory, depend on the spacetime dimension and call for a case by case discussion.
While the phenomenon of holographic degeneracy appears only for very specific dimensions, the conjectured bulk dual of the \textit{free} scalar CFT (Gaussian fixed point) is the same over all spacetime dimensions: Vasiliev higher-spin gravity with \textit{unbroken} higher-spin symmetries.

$\bullet$ At the kinematical level (\textit{i.e.} two-point functions) and in the large N limit,
the conjecture at the Gaussian fixed point
seems to be ensured by the Flato-Fronsdal theorem \cite{Flato:1978qz}, generalised
to any dimension in \cite{Vasiliev:2004cm}.
Strictly speaking, the conjecture implicitly involves a regularisation of the divergencies, which brings some subtleties even at the kinematical level \cite{Joung:2011xb}.

A constructive proof of the strong form of the conjecture at the Gaussian fixed point is particularly tantalising for two reasons:

$\bullet$ The validity of the strong form of the conjecture at the Gaussian fixed point, \textit{i.e.} for any N but at zero coupling, would be sufficient to formally define the anti-holographic dual of vector models with quartic interactions, for any values of N and of the coupling constant.

$\bullet$ The possibility of a direct proof at the Gaussian fixed point is quite plausible because both sides of the duality seem tractable, in the sense that they should be (almost) uniquely fixed by the huge symmetries. Indeed, unbroken higher spin symmetries appear to prevent genuine interactions on the boundary \cite{Maldacena:2011jn} and quantum corrections in the bulk \cite{Fradkin:1990kr}.\footnote{The group of higher-spin gauge symmetries is expected to be large enough to eliminate any non-trivial counter-term, as conjectured a while ago by Fradkin \cite{Fradkin:1990kr}.}

\noindent These features have brought some hope for a tractable proof of the AdS/CFT correspondence for the paradigmatic $O(N)$ model, maybe even beyond the semi-classical approximation (\textit{i.e.} large $N$ limit) and even outside the singlet sector.

\section{The free CFT}

Consider $n\,N$ free massless complex scalar fields $\phi^{i,a}\,$ with
$i=1,\ldots,N$ and $a=1,\dots,n$\,, living on the conformal boundary of $AdS_{d+1}$
(identified with the compactification of $\mathbb{R}^d$\,).
Later on, one will consider the large N limit while $n$ will always be kept fixed. 
The collection of $n\,N$ free massless scalar fields can also be seen as a set of $n$ vector multiplets
where each individual vector $\bm\phi^a=(\phi^{i,a})$ has $N$ components ($a$ is fixed).
The free action is the
quadratic functional: $S_{\rm free}[\bm\phi] =
    \int d^dx\ \delta_{ab}\,\partial_\mu\bm\phi^{*a}(x)\bm\cdot\partial^\mu\bm\phi^b(x)\,,$
where $\bm\phi^a_1\bm\cdot\bm\phi^b_2:=\delta_{ij}\,\phi_1^{i,a}\,\phi_2^{j,b}\,.$
This action is invariant under global $U(nN)$-transformations
of the vector multiplet (or $O(nN)$ if the scalars are real) and
under the conformal transformations. For example, under dilatation $\bm\phi^a$ transforms as
$\bm\phi^a(\lambda\,x)=\lambda^{-\Delta_{\bm\phi}}\,\bm\phi^a(x)\,,$ where
$\Delta_{\bm\phi}=\frac{d-2}2$ is the canonical dimension of the scalar fields.
The Euler-Lagrange equation is $\square\,\bm\phi(x)\approx 0$.
Equalities that are valid only on some mass shell will be denoted by a weak equality symbol $\approx$\,.

Since the theory is free, there are infinitely many conserved
currents, among which a special role is played by the following matrix-valued symmetric tensor fields:  \be
    J^{ab}_{\mu_1\cdots\mu_s}(x)\,=\,\Big(\frac{i}{2}\Big)^s
    \sum_{n=0}^s\ (-1)^n\,\binom{s}{n}\
    \partial_{(\mu_1}\dots\partial_{\mu_n}\bm\phi^{b}(x)
    \bm\cdot
    \partial_{\mu_{n+1}}\dots\partial_{\mu_s)}\bm\phi^{*a}(x)\,,
\label{currents}
\ee
where the round brackets stands for total symmetrisation (with unit weight).
The bilinear currents (\ref{currents}) are conserved and $U(N)$-singlet. Moreover, they take values in the adjoint\footnote{If the bosonic scalar fields are real, the currents are all $O(N)$-singlet 
and the ones of even (odd) rank take values in the (anti)symmetric representation of $\mathfrak{o}(n)\,$.} representation of $\mathfrak{u}(n)$, \textit{i.e.} $(J^{ab})^*=J^{ba}\,$.
Usually the conjecture is formulated for $n=1$ but the general case can also be considered \cite{Klebanov:2002ja}.
Notice that though the spin-$s$ conserved currents \eqref{currents} are not traceless, they are in one-to-one correspondence with the traceless ones. The explicit relations between them can be found in \cite{Bekaert:2010ky}. In any case, without loss of generality the conjecture can be stated in terms of the much simpler collection of bilinears \eqref{currents} which can be packed into a simple (bilocal) generating function:
\be
    J^{ab}(x,q)= \bm\phi^{*a}(x+\,i\,q/2)\bm\cdot\bm\phi^b(x-\,i\,q/2)=\sum_{s=0}^\infty\frac{1}{s!}\,
    J^{ab}_{\mu_1\cdots\mu_s}(x)\,q^{\mu_1}\cdots q^{\mu_s}\,.
    \label{genfct}
\ee
One can check that the equations of motion of $\bm\phi^{*a}$ and $\bm\phi^b$ imply the conservation condition,
\be
\left(\eta^{\mu\nu}\frac{\partial}{\partial{x^\mu}}\frac{\partial}{\partial{q^\nu}}\right)\,J^{ab}(x,q)\approx0\quad\Longleftrightarrow\quad\partial^\mu J^{ab}_{\mu\mu_1\cdots\mu_{s-1}}(x)\approx 0\,,\quad \forall s\geqslant 1\,.
\ee

\section{``Single-trace'' deformation of the free CFT}

By analogy with the standard AdS/CFT terminology, the $U(N)$-singlet currents (\ref{currents}) are usually called ``single-trace'' operators with a slight abuse of terminology since the scalar fields are in the fundamental representation rather than in the adjoint.
The generating functional $\mathcal{W}_0$ of the connected correlators of single-trace operators in the free CFT can be seen as the free energy of the free CFT deformed linearly by the latter operators, \textit{i.e.}
\be
    \exp\Big(-\,\mathcal{W}_0[h;N]\,\Big)=\int\mathcal{D}\bm\phi\,
    \exp\Big(\!-S_{\rm free}[\bm\phi]\,+\sum_{s=0}^\infty\frac{1}{s!}\,\int d^dx\, \big(\,h_{ab}^{\mu_1\cdots\mu_s}(x)J^{ab}_{\mu_1\cdots\mu_s}(x)+\mbox{c.c.}\,\big)\Big)\,,
    \label{connfree}
\ee where the $\mathfrak{u}(n)$-valued symmetric tensor fields
$h_{ab}^{\mu_1\cdots\mu_s}(x)$ are external sources. Formally, the
functional \eqref{connfree} is equal to \cite{Bekaert:2010ky} the one-loop
effective action $\mathcal{W}_0[h;N]=N
\mathcal{W}_0[h;1]=N\mbox{Tr}\log(\Box+\hat{H}_{ab})$ where
$\hat{H}_{ab}$ is the Hermitian operator whose Weyl symbol is
$h_{ab}(x,p)=\sum_{s=0}^\infty\frac{1}{s!}\,h_{ab}^{\mu_1\cdots\mu_s}(x)\,p_{\mu_1}\cdots
p_{\mu_s}$ \cite{Bekaert:2009}.

Following the Gubser-Klebanov-Polyakov-Witten (GKPW) prescription, the strong form of the higher-spin holography conjecture is that the bulk action ${\cal S}_0[{\cal H}]$ of the anti-holographic dual to the free $CFT_d$ corresponds to bosonic Vasiliev theory \cite{Vasiliev:2003ev} around $AdS_{d+1}$ and is such that\footnote{We assume the existence of a conventional action principle $\mathcal S_0$ for Vasiliev's unfolded equations (this issue is one of the major open question in higher-spin theory, but see \cite{Boulanger:2011dd} for an interesting exotic proposal) and we omit to write explicitly the gauge fixing terms and ghosts contributions.}
\be
    \exp\Big(-\,N\,\mathcal{W}_0[h;1]\,\Big)\,=\,\int\limits_{{\cal H}|_{\partial AdS}=h} {\cal D}\mathcal H\ \exp\Big(-\,N\,{\cal S}_0[{\cal H}]\,\Big)\,,
    \label{GKPWfree}
\ee
where ${\cal H}(x,z)$ denotes the $\mathfrak{u}(n)$-valued bulk tensor gauge fields dual to the single-trace operators $J(x)$ of scaling dimension $\Delta$
and the boundary condition ${\cal H}|_{\partial AdS}=h$ stands for the behaviour ${\cal H}(x,z)\sim z^{d-\Delta}h(x)$ in the limit $z\sim 0$.
To avoid confusion, one should repeat that although the boundary CFT is free, the duality is nevertheless non-trivial
because the bilinear operators actually couple to background sources and the boundary functional $\mathcal{W}_0$ is not quadratic, therefore the bulk dual theory is interacting.

The saddle point approximation of the equality \eqref{GKPWfree} implies that $\mathcal{W}_0[h;1]={\cal S}_0[{\cal H}(h)]$ in the large N limit where ${\cal H}(h)$ stands for the solution of the Vasiliev equations with the prescribed boundary condition.
Actually, the previous equality between the generating functional of connected correlators and the on-shell bulk action must be exact according to the strong form of the conjecture. In other words, the anti-holographic dual to the free CFT should not receive quantum corrections because the generating functional of connected correlators of singlet bilinears does not receive $1/N$ corrections (since it comes from a Gaussian integral). Since the bulk dual of the free CFT is conjectured to be a higher-spin theory in the bulk with unbroken gauge symmetries, this surprising property is plausible because the group of symmetries may be huge enough to eliminate any non-trivial counterterm.

\section{Two-point functions: the duality as Flato-Fronsdal theorem}\label{2pt}

At kinematical level (\textit{i.e.} at the level of two-point functions and in the large-N limit), the $AdS_{d+1}/CFT_d$ correspondence can be seen, from a group-theoretical perspective, as a mere intertwiner of representations of $\mathfrak{o}(d,2)$ \cite{Dobrev:1998md} where the latter algebra is either realised as $AdS_{d+1}$ isometries or as conformal transformations of $\partial AdS_{d+1}$.

The {maximal compact subalgebra} is $\mathfrak{o}(2)\oplus\mathfrak{o}(d)$ and corresponds to the {time translations} generated by the (conformal) {Hamiltonian} and to the $\mathfrak{o}(d)$-rotations,
both acting in a natural way on the boundary $\partial AdS_{d+1}\cong S^1\times S^{d-1}$.
Accordingly, the positive-energy unitary irreducible $\mathfrak{o}(d,2)$-modules denoted by
${\cal D}(\Delta,s)$ are caracterized by the energy (scaling dimension) $\Delta>0$ and by the spin $s\in\frac12{\mathbb N}$ of the ground state (conformal primary operator).
In particular, the elementary conformal scalar fields living on the boundary span the module ${\cal D}(\frac{d}2-1,0)$ called a scalar singleton.
The conserved traceless symmetric tensor fields of rank $s$ on the boundary span the module
${\cal D}(s+d-2,s)$. For instance, the composite bilinear fields \eqref{currents} span these modules, as follows from dimensional analysis.
Equivalently, the module ${\cal D}(s+d-2,s)$ is realized in terms of elementary bulk symmetric tensor gauge fields of rank $s$.

The generalized Flato-Fronsdal theorem states that the (anti) symmetric tensor product, denoted by $\vee$ (respectively, $\wedge$), of two scalar singletons ${\cal D}(\frac{d}2-1,0)$ decomposes as the infinite sum of all the modules ${\cal D}(s+d-2,s)$ with even (resp. odd) spin $s$ \cite{Vasiliev:2004cm}:
$$
{\cal D}\Big(\frac{d}2-1,0\Big)\vee\,{\cal D}\Big(\frac{d}2-1,0\Big)\,=\,\bigoplus\limits_{\mbox{s even}}^\infty{\cal D}(s+d-2,s)\,,\,\,\,\,{\cal D}\Big(\frac{d}2-1,0\Big)\wedge\,{\cal D}\Big(\frac{d}2-1,0\Big)\,=\,\bigoplus\limits_{\mbox{s odd}}^\infty{\cal D}(s+d-2,s)\,.
$$
This decomposition of the product of two scalar singletons underlies the expansion of the bilocal generating function
\eqref{genfct} with the properties explained above.
This theorem is the rigorous justification of the dictionary between boundary conserved currents and bulk gauge fields \cite{Konstein:2000bi,Mikhailov:2002bp,Sezgin:2002rt} and it ensures the validity of the conjecture at the kinematical level.

\section{``Double-trace'' deformation of the free CFT}\label{brok}

The scalar single-trace operator 
\be
{\cal O}(x):=\delta_{ab}J^{ab}(x)=\delta_{ab}\bm\phi^{*a}(x)\bm\cdot\bm\phi^b(x)
\label{single-trsc}
\ee has to be distinguished from the other bilinears for various reasons: Firstly, it corresponds to the degenerate case of a rank-zero ``current''. Secondly, it is an $U(nN)$ singlet. Thirdly, its anti-holographic dual is a scalar field ${\cal H}(x,z):=\delta^{ab}{\cal H}_{ab}(x,z)$ which for $2<d<4$ and $4<d<6$ lies in the range of mass allowing for two distinct quantizations (\textit{c.f.} Section \ref{Case}). Fourthly, the quartic vertex of the interacting $U(nN)$ model is equal to the square ${\cal O}^2$ of this operator. This vertex is usually called a ``double-trace deformation'' by analogy with the standard AdS/CFT jargon.
Most of the discussion below actually applies to the generic case of double-trace deformations \cite{Berkooz:2002ug,Witten:2001ua,Sever:2002fk}.

The generating functional $\mathcal{W}_\lambda$ of the connected correlation functions of $U(N)$-singlet bilinears in the interacting $U(nN)$ model is written in terms of the scalar source $\alpha(x):=\delta_{ab}h^{ab}(x)$ (the external sources carrying spin and flavor indices will be ommited in the sequel because, though they can be included, they do not play any role in the issue under discussion):
\be
    \exp\Big(-\mathcal{W}_\lambda[\alpha;N]\,\Big)\,=\,\int\mathcal{D}\bm\phi\,
    \exp\Big(-S_{\rm free}[\bm\phi]\,-G_\lambda[{\cal O},\alpha]\Big)\,,
    \label{connfull}
\ee
where the ``deformation'' is the functional
\be
G_\lambda[{\cal O},\alpha]=-\int d^dx\, \alpha(x){\cal O}(x)\,+\frac{\lambda}{2N}\int d^dx\, {\cal O}^2(x)
\label{defo}
\ee
so that $\mathcal{W}_{\lambda=0}[\alpha]$ corresponds to the free CFT deformed only linearly by the single-trace operator ${\cal O}(x)$
\be
    \exp\Big(-\mathcal{W}_0[\alpha;N]\,\Big)\,=\,\int\mathcal{D}\bm\phi\,
    \exp\Big(-S_{\rm free}[\bm\phi]\,+\int d^dx\, \alpha(x){\cal O}(x)\Big)\,.
    \label{connfree2}
\ee
The standard Hubbard-Stratonovich trick corresponds to the introduction of an auxiliary field $\sigma$ through a Gaussian integral
$$
    \exp\Big(-\mathcal{W}_\lambda[\alpha;N]\,\Big)\,=\,\int\mathcal{D}\bm\phi\,\int\mathcal{D}\sigma\,
    \exp\Big(-S_{\rm free}[\bm\phi]\,+\int d^dx\, \big(\alpha(x)+\sigma(x)\big){\cal O}(x)\,+\frac{N}{2\lambda}\int d^dx\, \sigma^2(x)\Big)\,.
$$
where one implicitly assume to be in the interacting case $\lambda\neq 0$. The integration over the dynamical field on the boundary can be performed by making use of \eqref{connfree2} which leads to
\be
    \exp\Big(-\mathcal{W}_\lambda[\alpha;N]\,\Big)\,=\,\int\mathcal{D}\sigma\,
    \exp\left(-\,N\,\left(\mathcal{W}_0[\alpha+\sigma;1]\,-\frac{1}{2\lambda}\int d^dx\, \sigma^2(x)\right)\,\right)\,.
    \label{HStransform}
\ee

Let us now examine a possible interpretation of the
Hubbard-Stratonovich transformation from the dual perspective. Let
us assume that there exists an anti-holographic dual to the free
CFT along the GKPW prescription \be
    \exp\Big(-N\mathcal{W}_0[\alpha;1]\,\Big)\,=\,\int\limits_{{\cal H}|_{\partial AdS}=\alpha} {\cal D}{\cal H}\ \exp\Big(-N{\cal S}_0[{\cal H}]\,\Big)\,.
    \label{GKPWfree2}
\ee
The relation \eqref{GKPWfree2} leads, upon differentiation with respect to $\alpha$, to the identification between the mean values of the single-trace operator $\cal O$ and of the boundary value $\beta=\delta {\cal S}_0[{\cal H}]/\delta \alpha$ of the conjugate of its dual field
\be
\frac{\delta\mathcal{W}_0[\alpha;1]}{\delta\alpha(x)}\,=\,-\,\langle {\cal O}(x)\rangle_\alpha\,=\,\,<\, \beta(x)>_\alpha\,,
\ee
where $\langle \,\,\rangle_\alpha$ and $<\, \,\,>_\alpha$ denote the mean values for the path integrals, respectively, in the boundary and in the bulk.
Inserting the GKPW prescription \eqref{GKPWfree2} inside the Hubbard-Stratonovich form of the CFT path integral \eqref{HStransform} leads to\footnote{Notice that the 1/N expansion corresponds to a loop expansion on both sides: for the Hubbard-Stratonovich field $\sigma$ on the boundary and for the bulk gauge fields $\cal H$ in the bulk.}
\be
    \exp\Big(-\mathcal{W}_\lambda[\alpha;N]\,\Big)\,=\,\int\mathcal{D}\sigma\int\limits_{{\cal H}|_{\partial AdS}=\alpha+\sigma} {\cal D}{\cal H}\ \exp\left(\,-\,N\,\left(\,{\cal S}_0[{\cal H}]\,-\frac{1}{2\lambda}\int d^dx\, \sigma^2(x)\,\right)\,\right)\,.
\ee
Performing the path integral over $\sigma$ should be equivalent to the relaxation of the boundary coundition on the bulk scalar field ${\cal H}$, leading to the following holographic relation
\be
\exp\Big(-\mathcal{W}_\lambda[\alpha;N]\,\Big)\,=\,\int{\cal D}{\cal H}\ \exp\Big(-\,N\,{\cal S}_\lambda[{\cal H},\alpha]\,\Big)\,,
\label{correspo}
\ee
where the action describing the bulk dual of the interacting CFT
\be
{\cal S}_\lambda[{\cal H},\alpha]\,=\,{\cal S}_0[{\cal H}]\,-\,\frac{1}{2\lambda}\int d^dx\, \left({\cal H}|_{\partial AdS}-\alpha\right)^2(x)
\label{bulkact}
\ee
is a deformation of the initial action ${\cal S}_0[{\cal H}]$ by a boundary term.
Although the holographic relation \eqref{correspo} follows from the mere combination of two conventional ingredients: the Hubbard-Stratonovich transformation and the GKPW prescription, it seems rather unconventional since \eqref{correspo} does not involve any explicit boundary condition for the bulk singlet scalar field \cite{Berkooz:2002ug}.
Actually, the boundary condition is somehow hidden in the extra boundary term added to the initial bulk action. Indeed, in the semi-classical approximation the bulk action should be evaluated on solutions where it is stationary
\be
\delta {\cal S}_\lambda[{\cal H}]\,=\,\int\limits_{AdS}\delta{\cal H}\,\frac{\delta {\cal S}_0[{\cal H}]}{\delta{\cal H}}\,+\, \frac{1}{\lambda}\int\limits_{\partial AdS} \delta{\cal H}|_{\partial AdS}\,\Big(\lambda\frac{\delta {\cal S}_0[{\cal H}]}{\delta{\cal H}|_{\partial AdS}}-{\cal H}|_{\partial AdS}+\alpha\Big)\approx 0\,.
\ee
So, on-shell, the undeformed equations of motion $\delta {\cal S}_0[{\cal H}]/\delta{\cal H}\approx 0$ should be satisfied but also the boundary condition
\be
{\cal H}|_{\partial AdS}-\lambda\frac{\delta {\cal S}_0[{\cal H}]}{\delta{\cal H}|_{\partial AdS}}\approx \alpha\,.
\label{bdycond}
\ee
This boundary condition is in agreement with Witten's prescription \cite{Witten:2001ua} for general multi-trace deformations, following which: (i) one should replace $\cal O$ by $-\beta$ in the CFT deformation $G[{\cal O}]$, here $G_\lambda[-\beta;\alpha]=\int d^dx\,\big( \alpha(x)\beta(x)+\frac{\lambda}{2}\,\beta^2(x)\big)$ due to \eqref{defo}, and (ii)
the boundary value of ${\cal H}$ is equal to the variation of this deformation with respect to $\beta$: ${\cal H}|_{\partial AdS}=\delta G_\lambda[-\beta;\alpha]/\delta\beta$, which gives here ${\cal H}|_{\partial AdS}= \alpha+\lambda\beta$, \textit{i.e.} \eqref{bdycond}. Notice that the boundary condition \eqref{bdycond} at $\lambda=0$ is the (``Dirichlet'' for $d>4$ versus ``Neumann'' for $2<d<4$) condition: ${\cal H}|_{\partial AdS}=\alpha$, while at $\lambda=\infty$ (in the IR for $2<d<4$) it is the conjugate (respectively, ``Neumann'' vs ``Dirichlet'') condition \cite{Klebanov:1999tb}: $\frac{\delta {\cal S}_0[{\cal H}]}{\delta{\cal H}|_{\partial AdS}}=\beta$.

The conclusion is that the bulk quantum path integral may not be
necessarily subject to any boundary prescription on the singlet
scalar field driving the deformation, the boundary condition only
arising in the semi-classical approximation \cite{Sever:2002fk}.
More precisely, the double-trace deformations in the CFT are dual
to quadratic boundary terms in the bulk that, in the
semi-classical approximation, are responsible for the boundary
conditions \cite{Berkooz:2002ug}. In the particular case of the
higher-spin conjecture, one is lead to the conclusion that the
exact duality (for any value of the coupling constant) follows
from its validity at the Gaussian fixed point (for vanishing
coupling constant) by adding a suitable boundary term and by
formally relaxing the boundary condition. In the large N limit,
this result is realized via the diagrammatic properties observed
in \cite{Giombi:2011ya}.

\section{Holographic degeneracy and dimensional analysis}\label{Case}

In low dimensions, the conjecture is particularly interesting because it provides a simple example of the phenomenon of holographic degeneracy
where two distinct renormalisation group fixed points can have the same bulk dual theory (though with distinct boundary conditions).

This phenomenon was observed in \cite{Klebanov:1999tb}
by considering a peculiarity of the $AdS_{d+1}/CFT_d$ dictionary for scalar fields/\,operators.
In the large N limit, the mass/\,dimension relation is $\Delta^2\,-\,d\,\Delta\,-\,m^2\,=\,0$ for spin zero. The phenomenon of holographic degeneracy appears when both roots,
$\Delta_\pm=\frac{d}2\pm[m^2+(\frac{d}{2})^2]^\frac12$,
are compatible with unitarity.
On the one hand, the requirement that the scaling dimensions $\Delta_\pm$ must be real is equivalent to the Breitenlohner-Freedman bound $m^2\geqslant -(\frac{\,d}{\,2})^2$ on the mass-square.
On the other hand, an admissible scaling dimension $\Delta$ must also satisfy the unitarity bound $\Delta\geqslant (d-2)/2$.
The ordering of the roots is such that $\Delta_+\geqslant \frac{d}2\geqslant\Delta_-$, thus the highest root $\Delta_+$ is always above the unitarity bound. The holographic degeneracy only happens for
$$ \Delta_+\geqslant \Delta_-\geqslant \frac{d-2}2 \quad\Longleftrightarrow\quad -\left(\frac{\,d}{\,2}\right)^2\leqslant m^2\leqslant 1-\left(\frac{\,d}{\,2}\right)^2\,.$$
From the boundary point of view, the highest (lowest) root corresponds to an IR (UV) attractive fixed point of the renormalisation group. From the bulk point of view, the highest (lowest) root corresponds to the standard (exotic) boundary condition \cite{Klebanov:1999tb}.

The composite scalar field \eqref{single-trsc} made of two conformal scalars and driving the quartic deformation
precisely fits into this scheme \cite{Klebanov:2002ja}. The bare scaling dimension of the composite operator is twice the canonical dimension of the elementary field:
$\Delta^{\mbox{free}}=2\Delta_{\bf \phi}=d-2$.
Therefore the mass-square of the dual bulk field is $m^2=-2(d-2)$. The other root is equal to $\Delta^{\mbox{int}}=2$ and this scaling dimension is anomalous ($\Delta^{\mbox{free}}\neq \Delta^{\mbox{int}}$) for $d\neq 4$. It should therefore correspond on the boundary to a non-trivial fixed point and to a change of boundary condition for the bulk scalar field in the higher-spin multiplet. On both sides (interacting CFT and AdS) the higher-spin symmetries (respectively, global and gauge) are broken at finite N.
These features are summarised in the table 1.
The phenomenon of holographic degeneracy
appears in the range $2\leqslant d\leqslant 6$ in the sense that both roots $\Delta_\pm$ are above the unitarity bound.\footnote{Notice that the Breitenlohner-Freedman bound is saturated in $d=4$ while the unitarity bound is saturated for $d=2$ and $d=6$.}
\begin{vchtable}[h]
\vchcaption{Scaling dimension of operator}
\begin{tabular}{|c|c|c|}
 \hline
$\Delta$ scalar & AdS boundary condition & CFT fixed point\\
\hline
$\Delta_+$ & Standard (``Dirichlet'') & IR \\\hline
$\Delta_-$ & Exotic (``Neumann'') & UV \\\hline
 \hline
$\Delta$ composite & Higher-spin symmetry & CFT fixed point\\
\hline
$\Delta^{\mbox{free}}=d-2$ & Unbroken  & Gaussian \\\hline
$\Delta^{\mbox{int}}=2\,\,\,\quad$ & Broken & Non-trivial \\\hline
\end{tabular}
\end{vchtable}

The renormalisation group behaviour of the $O(N)$ model in the large N limit
matches qualitatively with the previous $AdS_{d+1}/CFT_d$ dictionary.
Indeed, the beta function in the $O(N)$ model (in dimension $d\,$ greater than two\footnote{The dimension $d=2$ is exceptional because there the Gaussian fixed point saturates the unitarity bound and rather than a couple there is a line of fixed points. The higher-spin gravity is also quite different on $AdS_3$ because it is of Chern-Simons type. Recently a large litterature developped on the $AdS_3/CFT_2$ higher-spin holography but this case falls beyond the scope of this review so we do not attempt to consider it.}) is known explicitly at large N (see e.g. \cite{Moshe:2003xn}):
\begin{equation}
\beta(\tilde\lambda)\,=\,\Lambda\frac{d\tilde\lambda}{d\Lambda}\, =\,\frac{4-d}{\tilde\lambda_{\mbox{int}}}\,\,\tilde\lambda\,(\tilde\lambda-\tilde\lambda_{\mbox{int}})\,+\,{\cal O}(1/N)\,,
\label{beta}
\end{equation}
where $\Lambda$ is the momentum cutoff, $\tilde\lambda:=\Lambda^{d-4}\,g\,N$ denotes the (dimensionless) 't Hooft like coupling and $\tilde\lambda_{\mbox{int}}$ is a finite number which depends on the regularisation scheme but whose sign is always the same as $4-d$ (and so the front factor $(4-d)/\tilde\lambda_{\mbox{int}}$ is always positive).
The two fixed points are the Gaussian fixed point $\tilde\lambda=\tilde\lambda_{\mbox{free}}=0$ and the non-trivial fixed point
$\tilde\lambda=\tilde\lambda_{\mbox{int}}+{\cal O}(1/N)$.
By dimensional analysis, the scaling dimension $\Delta(\tilde\lambda)$ of the composite operator
in the vicinity of a fixed point $\tilde\lambda$ obeys to
$\Delta=\frac{d}{\,2}+\frac{\beta^\prime(\tilde\lambda)}2$, in agreement with
$\Delta^{\mbox{free}}=d-2$ and $\Delta^{\mbox{int}}=2$ in the large N limit.

The case by case discussion of the fixed points at large N goes as follows and is summarised in the table 2.
For instance, triviality in $d=4$ means that the fixed points should coincide and indeed one finds $\tilde\lambda_{\mbox{free}}=\tilde\lambda_{\mbox{int}}=0$ in agreement with $\Delta^{\mbox{free}}=\Delta^{\mbox{int}}=2$.
For $2<d<4$ both fixed points are admissible and \eqref{beta} indicates that the RG flow goes to the non-trivial Wilson-Fisher fixed point in the IR and to the Gaussian fixed point in the UV, in agreement with $\Delta_-=\Delta^{\mbox{free}}<\Delta^{\mbox{int}}=\Delta_+$.
The non-trivial fixed point formally exists in higher dimensions but it is not physically admissible because
it corresponds to an unstable potential ($\tilde\lambda_{\mbox{int}}<0$) when $d>4$. It even violates the unitarity bound ($\Delta^{\mbox{int}}=2<\frac{d-2}2$) when $d>6$.
Notice that \eqref{beta} indicates that the formal RG flow is opposite for $d>4$, it goes to the non-trivial fixed point in the UV and to the Gaussian fixed point in the IR, in agreement with $\Delta_-=\Delta^{\mbox{int}}<\Delta^{\mbox{free}}=\Delta_+$.\footnote{Exactly the same renormalisation group behaviour holds for non-relativistic bosons and fermions in $d-2$ spatial dimensions at zero chemical potential \cite{Nikolic:2007zz} with the important difference that, for fermions, the non-trivial fixed point is also admissible in the window $2<d-2<4$ and it corresponds to the unitary Fermi gas. This discussion underlies the holographic proposal \cite{Bekaert:2011cu}.}

\begin{vchtable}[h]
\vchcaption{Spacetime dimension versus scaling dimension}
\begin{tabular}{|c|c|c|c|}
 \hline
$d$ & $\Delta_-$ & $\Delta_+$ & Property\\
\hline\hline
$2$ & $\Delta^{\mbox{free}}=0$ & $\Delta^{\mbox{int}}=2$ & saturation of unitarity bound \\
& & & (line of fixed pts)\\\hline
$3$ & $\Delta^{\mbox{free}}=1$ & $\Delta^{\mbox{int}}=2$ & pair of admissible fixed pts \\
 & &  & (asymptotic freedom) \\\hline
$4$ & $\Delta^{\mbox{free}}=\Delta^{\mbox{int}}=2$ & $\Delta^{\mbox{free}}=\Delta^{\mbox{int}}=2$ & fusion of fixed pts\\
&&& (triviality)\\\hline
$5,6$ & ($\Delta^{\mbox{int}}=2$) & $\Delta^{\mbox{free}}=3,4$ & only single admissible fixed pt \\
 & &  & (unstable interactions) \\
\hline
$\geqslant 7$ & ($\Delta^{\mbox{int}}=2$) & $\Delta^{\mbox{free}}\geqslant 5$ & only single admissible fixed pt\\
&&&(non-unitary interactions)\\\hline
\end{tabular}
\end{vchtable}

\section*{Acknowledgments}

X.B. is grateful to S. Moroz for useful discussions and warmly thanks the organisers of the XVII European Workshop on String Theory for this enjoyable meeting
and for the opportunity to present a talk and contribute to the proceedings.

%

\begin{thebibliography}{[1]}

\bibitem{Flato:1978qz}
  M.~Flato and C.~Fronsdal,
  Lett.\ Math.\ Phys.\  {\bf 2} (1978) 421.

\bibitem{Fronsdal:1978vb}
  C.~Fronsdal,
  Phys.\ Rev.\  D {\bf 20} (1979) 848.

\bibitem{Sundborg:2000wp}
  B.~Sundborg,
  Nucl.\ Phys.\ Proc.\ Suppl.\  {\bf 102} (2001) 113
  [arXiv:hep-th/0103247];
  E.~Witten, ``Spacetime reconstruction'' 
talk given at `J.H. Schwarz 60th Birthday Conference' (Cal Tech, November 2001).

\bibitem{Konstein:2000bi}
  S.~E.~Konstein, M.~A.~Vasiliev and V.~N.~Zaikin,
  JHEP {\bf 0012} (2000) 018
  [hep-th/0010239].

\bibitem{Mikhailov:2002bp}
  A.~Mikhailov,
  arXiv:hep-th/0201019.

\bibitem{Sezgin:2002rt}
  E.~Sezgin and P.~Sundell,
  Nucl.\ Phys.\  B {\bf 644}, 303 (2002)
  [arXiv:hep-th/0205131].

\bibitem{Klebanov:2002ja}
  I.~R.~Klebanov and A.~M.~Polyakov,
  Phys.\ Lett.\  B {\bf 550} (2002) 213
  [arXiv:hep-th/0210114].

\bibitem{Sezgin:2003pt}
  E.~Sezgin and P.~Sundell,
  JHEP {\bf 0507}, 044 (2005)
  [arXiv:hep-th/0305040].

\bibitem{Vasiliev:2003ev}
  M.~A.~Vasiliev,
  Phys.\ Lett.\  B {\bf 567} (2003) 139
  [arXiv:hep-th/0304049].\\
  For a review, see \textit{e.g.} X.~Bekaert, S.~Cnockaert, C.~Iazeolla and M.~A.~Vasiliev,
  arXiv:hep-th/0503128.



\bibitem{Giombi:2009wh}
  S.~Giombi and X.~Yin,
  JHEP {\bf 1009} (2010) 115
  [arXiv:0912.3462 [hep-th]];
  JHEP {\bf 1104} (2011) 086
  [arXiv:1004.3736 [hep-th]].

\bibitem{Aharony:2011jz}
  O.~Aharony, G.~Gur-Ari and R.~Yacoby,
  arXiv:1110.4382 [hep-th];
  S.~Giombi, S.~Minwalla, S.~Prakash, S.~P.~Trivedi, S.~R.~Wadia and X.~Yin,
  arXiv:1110.4386 [hep-th].

\bibitem{Anninos:2011ui}
  D.~Anninos, T.~Hartman and A.~Strominger,
  arXiv:1108.5735 [hep-th].

\bibitem{Jevicki:2011ss}
	S.~R.~Das and A.~Jevicki,
  Phys.\ Rev.\  D {\bf 68} (2003) 044011
  [arXiv:hep-th/0304093];
    R.~d.~M.~Koch, A.~Jevicki, K.~Jin and J.~P.~Rodrigues,
  Phys.\ Rev.\  D {\bf 83}, 025006 (2011)
  [arXiv:1008.0633 [hep-th]];
  A.~Jevicki, K.~Jin and Q.~Ye,
  J.\ Phys.\ A  {\bf 44}, 465402 (2011)
  [arXiv:1106.3983 [hep-th]].

\bibitem{Douglas:2010rc}
  M.~R.~Douglas, L.~Mazzucato and S.~S.~Razamat,
  Phys.\ Rev.\  D {\bf 83}, 071701 (2011)
  [arXiv:1011.4926 [hep-th]].

\bibitem{Bekaert:2011cu}
  X.~Bekaert, E.~Meunier and S.~Moroz,
  arXiv:1111.1082 [hep-th];
  arXiv:1111.3656 [hep-th].

\bibitem{Vasiliev:2004cm}
  M.~A.~Vasiliev,
  JHEP {\bf 0412} (2004) 046
  [arXiv:hep-th/0404124].

\bibitem{Joung:2011xb}
  E.~Joung and J.~Mourad,
  arXiv:1112.5620 [hep-th].

\bibitem{Maldacena:2011jn}
  J.~Maldacena and A.~Zhiboedov,
  arXiv:1112.1016 [hep-th].

\bibitem{Fradkin:1990kr}
E.~S. Fradkin, ``The problem of unification of all interactions and
  self-consistency,'' preprint Lebedev 90-0193, talk given at `Dirac Medal for
  1988' (Trieste, April 1989).

\bibitem{Bekaert:2010ky}
  X.~Bekaert, E.~Joung and J.~Mourad,
  JHEP {\bf 1102} (2011) 048
  [arXiv:1012.2103 [hep-th]].

\bibitem{Bekaert:2009}
  X.~Bekaert, E.~Joung and J.~Mourad,
  JHEP {\bf 0905}, 126 (2009)  [arXiv:0903.3338 [hep-th]].

\bibitem{Boulanger:2011dd}
  N.~Boulanger and P.~Sundell,
  J.\ Phys.\ A  {\bf 44} (2011) 495402
  [arXiv:1102.2219 [hep-th]].

\bibitem{Dobrev:1998md}
  V.~K.~Dobrev,
  Nucl.\ Phys.\  B {\bf 553}, 559 (1999)
  [arXiv:hep-th/9812194].

\bibitem{Klebanov:1999tb}
  I.~R.~Klebanov and E.~Witten,
  Nucl.\ Phys.\  B {\bf 556} (1999) 89
  [arXiv:hep-th/9905104].

\bibitem{Gubser:2002vv}
  S.~S.~Gubser and I.~R.~Klebanov,
  Nucl.\ Phys.\  B {\bf 656} (2003) 23
  [arXiv:hep-th/0212138].

\bibitem{Berkooz:2002ug}
  M.~Berkooz, A.~Sever and A.~Shomer,
  JHEP {\bf 0205} (2002) 034
  [arXiv:hep-th/0112264].

\bibitem{Witten:2001ua}
  E.~Witten,
  arXiv:hep-th/0112258.

\bibitem{Sever:2002fk}
  A.~Sever and A.~Shomer,
  JHEP {\bf 0207} (2002) 027
  [arXiv:hep-th/0203168].

\bibitem{Giombi:2011ya}
  S.~Giombi and X.~Yin,
  arXiv:1105.4011 [hep-th].

\bibitem{Moshe:2003xn}
  M.~Moshe and J.~Zinn-Justin,
  Phys.\ Rept.\  {\bf 385} (2003) 69
  [arXiv:hep-th/0306133] Section 2.

\bibitem{Nikolic:2007zz}
  P.~Nikolic and S.~Sachdev,
  Phys.\ Rev.\  A {\bf 75} (2007) 033608
  [arXiv:cond-mat/0609106].

\end{thebibliography}
%

\end{document}